\documentclass[a4paper,10pt]{article}
\usepackage[utf8x]{inputenc}
\usepackage{amsmath}
\usepackage{amsfonts}
\usepackage{amssymb}

\title{Multiverse in the Third Quantized  Horava-Lifshits Theory of Gravity }
\author{Mir Faizal\\ faizal.mir@durham.ac.uk  \\
Department of Mathematics,\\ University of  Durham, Durham, UK  \\
 }

\begin{document}

\maketitle

\begin{abstract}
In this paper we  analyze the third quantization of Horava-Lifshits theory of gravity without detail balance. 
We  show that the Wheeler-DeWitt equation for 
 Horava-Lifshits theory of gravity in minisuperspace approximation becomes the equation for time-dependent harmonic oscillator. After interpreting 
the scaling factor as the time, we are able to derive the third quantized wavefunction for multiverse. We  also show in third quantized formalism it 
is possible that the universe can form from nothing. After  we go on to analyze  the effect of introducing interactions in the Wheeler-DeWitt equation.
We see how this model of interacting universes can be used to explain  baryogenesis  with violation of baryon number conservation in the multiverse. We also 
analyze  how this model can 
possibly explain the present value of the cosmological constant.  
Finally we analyze    the possibility of the multiverse being formed from perturbations around a false vacuum and its decay to a true vacuum. 
\end{abstract}
\maketitle

\section{Introduction}
Wheeler-DeWitt equation is basically the Schroedinger equation for the universe. Just like in  first quantized quantum mechanics 
there is no way to account for the creation and annihilation of particles, there is no way to account for the topology of the universe to change 
by the creation and annihilation of universes in the second quantized Wheeler-DeWitt equation \cite{1b}. However it is well known that we 
interpret the Schroedinger wave equation
as a classical field equation, in the second quantized formalism. We account for the creation and annihilation of particles by  modifying 
the original Schroedinger wave equation by the addition of  non-linear terms to it. 
In analogy with second quantization of the Schroedinger  wave equation, in third quantization the Wheeler-DeWitt equation is viewed as a 
classical field equation generated by a certain classical action. The classical field theory thus obtained is then third quantized.  
The addition of non-linear terms to the Wheeler-DeWitt equation can account for the topology change generated by the creation and annihilation 
of universes. Thus third quantization naturally gives  rise to a theory of many  universes which is  called the multiverse \cite{ib}.   

The third quantization  has been 
 discussed implicitly in Refs. \cite{2b,3b}  and explicitly in Refs.  \cite{2ba,3ba}. The modification of Wheeler-DeWitt equation by the addition of 
non-linear terms and the third quantization of the resultant theory was formally analyzed  in Ref. \cite{sg}. Third quantization of   Brans-Dicke 
theories  \cite{i} and Kaluza-Klein theories  \cite{ia} has also been done.
Coherent states for the multiverse  have also been  studied in the third quantized formalism   \cite{81, 9}.
    
The existence of the multiverse was first proposed for giving a consistent explanation to the measurement problem in 
quantum mechanics and in this context it was called  the Everett’s
many-world interpretation of the quantum theory
\cite{1a}. This idea  also appeared in the   landscape of the string theory \cite{3a, 4a}. 
In string theory 
 the number of different
vacuum states is  estimated to be  around $10^{500}$ \cite{1h}.
 It is suspected that all these different vacuum states could be real vacuum states of different universes \cite{1w}.  
The transition from one vacuum state to another has been 
 used as an explanation for inflation in the chaotic inflationary multiverse \cite{2a}. 
 
In this paper we will study the third 
quantization of the Horava-Lifshits theory of gravity  \cite{3, 4}.   Horava-Lifshits theory of gravity is a ultraviolet completion 
of gravity such that it reproduces general relativity in the infrared limit. Horava-Lifshits theory of gravity is 
motivated by the fact that the addition of higher order  curvature invariants 
leads to a renormalizable theory of gravity  \cite{1}, but at the cost of ghost states which spoil the unitarity of the theory \cite{2}. 
By breaking the Lorentz invariance of the theory and having  different Lipschitz scaling for the space and time, it is possible to add higher order spatial 
derivatives without adding any higher order temporal derivative. Thus it is possible to avoid ghost states but at the cost of breaking Lorentz invariance, 
in a renormalizable theory of gravity. 
As the theory uses the concept of  Lipschitz scaling from solid state physics, it is generally called Horava-Lifshits theory of gravity. 

 The original Horava-Lifshits theory of gravity is based on two assumptions which are called the detailed balance and the projectability \cite{5}. 
The projectability condition is related to the invariance with respect to time reparametrization and therefore to the Wheeler-DeWitt equation. 
  However as both these  assumptions of Horava-Lifshits theory of gravity are made just for simplification, 
it is possible to discuss a more generalized 
Horava-Lifshits theory of gravity  without these assumptions.  In fact  a generalization to the Horava-Lifshits theory of gravity without detailed 
balancing has already been done \cite{7, 8}.    
 In this paper we will analyze   the Wheeler-DeWitt equation for  generalized  Horava-Lifshits theory of gravity
without detailed balance in third quantized formalism. We will then study multiverse in this theory. 

\section{Wheeler-DeWitt Equation For  Horava-Lifshits Theory of Gravity}
For Einstein gravity, the 
Friedman-Robertson-Walker metric is given by 
\begin{equation}
ds^{2}=-N^{2}dt^{2}+a^{2}\left(  t\right)  d\Omega_{3}^{2}.
\end{equation}
where $d\Omega_{3}^{2}$ is the usual line element on the three sphere and $N$ is the
lapse function and here we have set $k=1$. In this background, we have $R_{ij} = 2 \gamma_{ij}/a^{2} $ and $R = 6/a^{2}$. 
Now following Horava  \cite{3}, we take the space and time to  exhibit the following Lipschitz scale
invariance 
\begin{eqnarray}
t&\rightarrow& \ell^{3}t, \nonumber \\ x^{i}&\rightarrow& \ell x^{i}.
\end{eqnarray}
With this Lipschitz scaling the four dimensional diffeomorphism
invariance   of the theory is  explicitly broken  and this  intern allows us to consider  different kinds of  kinetic and potential terms. 
The  
kinetic term  is taken to be quadratic in time derivatives of
the metric but the  potential term  contains  high-order space
derivatives. Now if $\mathcal{L}_{K}$ is the Kinetic term and $\mathcal{L}_{P}$ is the potential term, 
then the total action for Horava-Lifshits gravity can be written as 
\begin{equation}
S=\int dtd^{3}x\left(  \mathcal{L}_{K}-\mathcal{L}_{P}\right).
\end{equation}
Here the Kinetic term is given by
\begin{equation}
\mathcal{L}_{K}=N\sqrt{g}\frac{2}{\kappa^{2}}\left(  K^{ij}K_{ij}-\lambda
K^{2}\right),
\end{equation}
where $K_{ij}$ is the extrinsic curvature, which is defined by
\begin{equation}
K_{ij}=-\frac{1}{2N}\dot{g}_{ij}+\nabla_{i}N_{j}+\nabla_{j}N_{i} ,
\end{equation}
and $K=K^{ij}g_{ij}$ is its trace. There is no contribution  from the shift function to the extrinsic curvature in our model as
 the Friedman-Robertson-Walker metric
does not contain any contribution from the shift function.  A general potential term without detail balancing can be written as \cite{7} 
\begin{eqnarray}
\mathcal{L}_{P}&=& N\sqrt{g}[  g_{0}\zeta^{6}+g_{1}\zeta^{4}
R+g_{2}\zeta^{2}R^{2} +g_{3}\zeta^{2}R^{ij}R_{ij}+g_{4}R^{3}\nonumber \\ &&
  +g_{5}R\left(  R^{ij}R_{ij}\right)   +g_{6}R_{j}^{i}R_{k}^{j}R_{i}
^{k}+g_{7}R\nabla^{2}R\nonumber \\ &&  +g_{8}\nabla_{i}R_{jk}\nabla^{i}R^{jk}],
\end{eqnarray}
where the couplings $g_{a}  $ are all dimensionless. We can now use projectability to set $N=1$. 
 Plugging the action  for this generalized  Horava-Lifshits theory of gravity
without detailed balance into the Friedman-Robertson-Walker metric, and following the usual procedure to obtain the  Hamiltonian constraint, we get  
\begin{equation}
[\pi_{a}^{2}-\omega^2(a)]=0, 
\end{equation}
where 
\begin{eqnarray}
 \omega^2(a)  &=& - \frac{\left(  3\lambda-1\right)  }{\kappa^{2}}24\pi^{4}%
 \left[- g_{0}\zeta^{6}a^{4} +6\zeta^{4}a^{2}  
-6\zeta^{2}(6g_{2}+ 
g_{3})\right. \nonumber \\ && \,\,\,\,\,\,\,\,\, \left. -6(36g_{4}+6g_{5}+g_{6})a^{-2}  \right].
\end{eqnarray}
Now we promote the canonical momentum $\pi_{a}$ to an operator, and so we have 
$\pi_{a}=-i\partial_{a}$.
Thus the Wheeler-DeWitt equation corresponding to this classical Hamiltonian constraint is given by
\begin{equation}
\left[\frac{\partial^{2}}{\partial a^{2}}+ \omega^2(a)  \right]\phi [a] =0. \label{a}
\end{equation}
In this section we analyzed  the Wheeler-DeWitt equation  for Horava-Lifshits theory of gravity without detail balancing.
In the conventional second quantized formalism this
Wheeler-DeWitt equation would be interpreted  as the Schroedinger equation for our universe. 
However we will see in the next section that in order to study multiverse, it is best to interpret  this equation as a 
classical field equation and then third quantize it. 
\section{Third Quantization}

Wheeler-DeWitt equation when interpreted   as a quantum mechanical Schroedinger equation  can describe the quantum state of a
single universe.  However, just like the  first quantized Schroedinger equation is not compatible   with creation and annihilation of particles, 
the second-quantized Wheeler-DeWitt equation  is not compatible with topological changes associated with the
creation or annihilation of baby universes. These topology changing processes can be incorporated naturally by  using a third quantized formalism in analogy to 
use of second quantized formalism to account for the creation and annihilation of particles. Thus third quantization is an ideal 
formalism to study the multiverse. 

We interpret  Eq. $(\ref{a})$, 
as a classical field equation of a classical 
field $\phi(a)$, whose classical  action is given by 
\begin{equation}
\mathcal{S} [\phi (a)] = \frac{1}{2}
\int  da \left( \left({\frac{\partial \phi}{\partial a}}\right)^2 
- \omega^2(a) \phi^2 \right). \label{t}
\end{equation}
 Now obviously the variation of the third quantized action $\mathcal{S}[\phi (a)] $ given by Eq. $(\ref{t})$,
 will lead to the Wheeler-DeWitt equation $(\ref{a})$. 
If we  interpret  the scaling factor  $a$ as the time variable \cite{l1} then
 we can calculate the momentum conjugate to $\phi(a)$ as 
\begin{equation}
p_{\phi}=\frac{\delta \mathcal{S} [\phi (a)]  }
{\delta \dot{\phi}} = \dot{\phi},
\end{equation}
where 
\begin{equation}
 \dot{\phi} = \frac{\partial \phi}{\partial a }. 
\end{equation}
This the third quantized  Hamiltonian obtained by 
the Legendre transformation,   can be written as 
\begin{equation}
\mathcal{H} = \frac{1}{2} p_{\phi}^2 + \frac{\omega^2(a)}{2}
\phi^2 . \label{c}
\end{equation}
This  Hamiltonian given by Eq. $(\ref{c})$  is the  Hamiltonian for the harmonic oscillator with
time-dependent frequency $\omega(a)$. The solution to the classical equation of motion for this oscillator is given by 
\begin{equation}
    \phi(a)=\rho(a)\left[c_1 \cos\gamma(a)+c_2 
    \sin\gamma(a)\right],
\end{equation}
where the constants $c_1$ and $c_2$ are  determined by the boundary conditions used. If we use the boundary condition $\phi_1(a_1) = \phi_1$ and
 $\phi_2 (a_2) = \phi_2$, then we have 
\begin{eqnarray}
 c_1&=&\frac{1}{\sin(\gamma_2-\gamma_1)}\left(\frac{\phi_1}{\rho_1}
\sin\gamma_2-\frac{\phi_2}{\rho_2}\sin\gamma_1\right),\nonumber \\
c_2&=&\frac{1}{\sin(\gamma_2-\gamma_1)}\left(\frac{\phi_2}{\rho_2}
\cos\gamma_1-\frac{\phi_1}{\rho_1}\cos\gamma_2\right).
\end{eqnarray}

As the scaling factor $a$ is  interpreted as the time in this formalism, we can  
write the  third quantized
Schroedinger equation for this time-dependent harmonic oscillator as follows, 
\begin{equation}
\mathcal{H} \Phi= i  \frac{\partial}{\partial a}
\Phi.  \label{q}
\end{equation}
Now we can easily calculate the propagator for this  harmonic oscillator with time-dependent frequency, 
\begin{eqnarray}
G(  \phi_2,  \phi_1)&=&\frac{1}{2^n
n!}\left[\frac{1}{2i\pi  \rho_1\rho_2\sin(\gamma_2-\gamma_1)}\right]^{1/2}
\exp \left[\frac{   i \phi_2^2}{4}\left(\frac{\dot{\rho}_2}{\rho_2}\right) \right.
 \nonumber\\&&-\left.
    \frac{   i \phi_1^2}{4}\left(\frac{\dot{\rho}_1}{\rho_1}\right)\right]
    \exp\left[-\frac{1}{2  }\left(\frac{  \phi_2^2}{\rho_2^2}
    +\frac{  \phi_1^2}{\rho_1^2}\right)\right]\sum H_{n}\left[\frac{  \phi_2}{\rho_2}\right]
   \nonumber\\&&\times H_{n}\left[\frac{  \phi_1}{\rho_1}\right]
\exp\left(-i\gamma\left(n+\frac{1}{2}\right)\right),
\end{eqnarray}
where $H_{n}$ is Hermite polynomials. 
If $\Phi_n(\phi, a) $ is the amplitude to detect $n$ universes with the scalar factor $a$ in the multiverse, then we have 
\begin{eqnarray}
 \Phi_{n}(\phi,a)&=&\exp[i\alpha_{n}(a)]\left[\frac{1}{\pi^{1/2}n!2^{n}\rho}\right]^{1/2}
 \exp\left[\frac{i }{2}\left(\frac{\dot{\rho}}{\rho}+\frac{i}{\rho^{2}}\right)\phi^{2}\right]
 \nonumber\\&&\times H_{n}\left[\frac{\phi}{\rho}\right],
\end{eqnarray}
where  the phase functions $\alpha_{n}(a)$ are described by
\begin{eqnarray}
\alpha_{n}(a)=-\left(n+\frac{1}{2}\right)\int_{0}^{a}\frac{1}{\rho^{2}}da'.
\end{eqnarray}
 Thus the wavefunction for the full multiverse is  given by 
\begin{equation}
 \Phi  =  \sum_n C_n \Phi_n.
\end{equation}
This wavefunction can be obtained by a unitary transformation from the wavefunction of the usual 
 harmonic oscillator with time-independent frequency \cite{l2,l4}.
 Thus if $\tilde{\Phi}$ is the wavefunction of the time-independent oscillator then $\Phi$ can be written as 
\begin{equation}
 \Phi = \frac{1}{\sqrt{\rho}} U^{\dagger} \tilde{\Phi},
\end{equation}
where $1/\sqrt{\rho}$ is a normalization factor and $U$ is given by 
\begin{equation}
 U = \exp   \left(  \frac{i \dot{\rho}\phi^2}{2\rho}\right). 
\end{equation}
Thus in this representation  the Hamiltonian $\mathcal{H}$ can be written  as 
\begin{equation}
 \mathcal{H} = U^{\dagger}\mathcal{H}_0 U.
\end{equation}
This is the Hamiltonian for  Harmonic oscillator with  time-independent frequency which is denoted here by $\omega_0$, 
\begin{equation}
 \mathcal{H}_0 =\frac{1}{2} p_{0\phi}^2 + \frac{\omega_0^2}{2}
\phi_0^2.
\end{equation}
Now if  $b_0$ and $b^{\dagger}_0$ are  the usual creation and annihilation operators given by 
\begin{eqnarray}
 b_0 &=& \sqrt{\frac{\omega_0}{2}}\left( \phi_0 +  \frac{i}{\omega_0} \phi_0\right), \nonumber \\ 
b^{\dagger}_0 &=& \sqrt{\frac{\omega_0}{2}}\left( \phi_0 - 
 \frac{i}{\omega_0} \phi_0\right).
\end{eqnarray}
These creation and annihilation operators satisfy 
\begin{eqnarray}
 \left[b_0,b_0^{\dagger} \right] = 1, &&
  \left[b_0,b_0\right] = 0,\nonumber \\  
 \left[b_0^{\dagger},b_0^{\dagger}\right] = 0. &&
\end{eqnarray}
We can now define a vacuum state as a state annihilated by $b_0$,  
\begin{equation}
 b_0 |0\rangle =0.
\end{equation}
Here the vacuum state represents a state of nothing, which has no space, time or matter fields in it. 
The action of $b_0^{\dagger}$ on the vacuum state 
creates a universe, i.e., spacetime. However this vacuum is not uniquely defined as this expression for the creation and annihilation operators 
is valid only  for 
$\omega =\omega_0$. The time-dependent annihilation and creation operators are given  by \cite{l5},
\begin{eqnarray}
b(a) &=& \mu(a) b_0 + \nu(a) b_0^\dag , \nonumber  \\ b^\dag(a) &=& \mu^*(a)
b_0^\dag + \nu^*(a) b_0,
\end{eqnarray}
where 
\begin{eqnarray}
\mu(a) &=& \frac{1}{2} \left( \frac{1}{\rho(a)} + \rho(a) - i
\dot{\rho}(a) \right) , \\ \nu(a) &=& \frac{1}{2} \left(
\frac{1}{\rho(a)} - \rho(a) - i \dot{\rho}(a) \right) ,
\end{eqnarray}
with, $|\mu|^2 - |\nu|^2 = 1$. Thus if we start from a vacuum $|0\rangle$, the number of universes formed from nothing at time $a$ will be given by 
\begin{equation}
 N = \langle 0| b^{\dagger}(a) b (a)  |0\rangle = |\nu(a)|^2. 
\end{equation}
Thus there is a finite probability for the formation of universes from nothing. However as we have third quantized the free
 Wheeler-DeWitt equation these universes formed from nothing will not interact with each other. 
In the next section we will analyze  a system of interacting universes by modifying the Wheeler-DeWitt equation in a non-linear way. 
\section{Interactions}
In the previous sections we have seen that the Wheeler-DeWitt equation is analogous to  Schroedinger wave equation,
 in the sense it represents the quantum state of a 
single universe. We have also seen that if we third quantize this equation then it represents the quantum state of an ensemble of non-interacting universes. 
This is still not enough to account for topology change. To obtain a theory consistent with topology change we need to include interaction terms. 
 So in this section we will analyze  the effect of introducing  interactions terms in the third quantized action for the Wheeler-DeWitt equation. We modify
the third quantized action given by Eq. $(\ref{t})$ as
\begin{equation}
 \mathcal{S}_{T} [\phi (a)] =  \mathcal{S} [\phi (a)] 
+  \mathcal{S}_I[\phi (a)].
\end{equation}
Now if we go to imaginary time $a \to ia$, then we can write the Euclidean partition function for this action as 
\begin{equation}
 Z = \int D\phi \,\exp - \mathcal{S} [\phi ]_{ET},
\end{equation}
where $\mathcal{S}[\phi]_{ET}$ is the Euclidean version of $\mathcal{S} [\phi]_{T}$ obtained by letting $a \to i a$. We can now calculate 
the $n$-point functions for any interacting term  by the usual methods. 

A simple cubic interaction term given by 
\begin{equation}
  \mathcal{S}_{EI} = \frac{\lambda }{6}
\int  da  \phi^3(a),
\end{equation}
will generate a three-point function given by 
\begin{equation}
 G(a_1, a_2, a_3)_E = - \lambda \int da_0 G(a_1, a_0)_E 
G(a_2,a_0)_E G(a_3,a_0)_E, 
\end{equation}
where $G(a_1, a_0)_E, G(a_2,a_0)_E$ and $ G(a_3,a_0)_E$ are the Euclidean Green's functions obtained from  
  $ \mathcal{S}_E$, which is the Euclidean version of $ \mathcal{S}$.  
This process represents the splitting of  a universe  $U_1$ with scaling factor $a_1$ into two universes, $U_2$ and $U_3$ with
 scalar factors $a_2$ and  $a_3$, respectively. 
Now if we represent matter and gauge fields collectively by $\chi$ and include the contribution  from $\chi$ in our formalism,
then $\phi$ would also depend on $\chi$, 
$\phi = \phi (a, \chi)$. 
Thus the three-point function in reality would represents the splitting of the universe $U_1$ with scaling factor $a_1$ and  matter and gauge  
field content  $\chi_1$, into two universes  $U_2$ and $U_3$ with
 scalar factors $a_2, a_3$ and their matter and gauge field contents   $\chi_2, \chi_3$, respectively. 
Now if the total number of baryons and anti-baryons in universe $U_1$ are $n_1 $ and $m_1$, respectively and  the universe $U_1$
 has formed from nothing without violating the baryon number conservation, then we have 
\begin{equation}
 Bn_1 - Bm_1 =0, 
\end{equation}
where $Bn_1$ and $Bn_2$ represent the total baryon number of the baryons $n_1$ and anti-baryons $n_2$, respectively. 
If the total number of baryons and anti-baryons in the universes $U_2$ and $U_3$ are $n_2,m_2$ and $n_3, m_3$, respectively, the baryon number 
conservation implies that
\begin{equation}
 Bn_2 + Bn_3 - Bm_2 -Bm_3 =0. 
\end{equation}
However the  baryon number in the universes $U_2$ or $U_3$ need not be separately conserved   
\begin{eqnarray}
 Bn_2 - Bm_2 &\neq& 0,\nonumber \\ 
 Bn_3 - Bm_3 &\neq&0.
\end{eqnarray}
Thus after splitting of the universe $U_1$,  the universe  $U_2$ can have more matter than anti-matter if the universe $U_3$ has more 
anti-matter than matter. This will not violate the baryon number conservation as the total 
baryon number of both the universes is still collectively conserved. This can possibly explain the domination of 
matter over anti-matter in our universe without violating baryon number conservation \cite{4z}.
  
In fact we could also study other forms of potential e.g.., $\phi^4$ interaction term. In this case we would generate a four-point function given by
\begin{equation}
 G(a_1, a_2, a_3, a_4)_E = - \lambda \int da_0 G(a_1, a_0)_E 
G(a_2,a_0)_E G(a_3,a_0)G(a_4,a_0)_E. \label{p}
\end{equation}
 This can represent collision of two universes  with scalar factors $a_1$ and $a_2$ to form two new universes with scalar factor $a_3$ and $a_4$, respectively.
 In fact we can view big bang in this model as the collision of two earlier universes 
to form our present day universe. Again by introducing matter and gauge fields and repeating the above argument we can show that even in this model one universe 
can have more matter than anti-matter without violating baryon number conservation. 

The process given by Eq. $(\ref{p})$ can also be interpreted as the  splitting of one  universe into three 
 distinct regions of spacetime. It may be interesting to note that in the formation of black holes an initial 
spacetime gives rise to a black hole, a white hole 
and another distinct region of spacetime. So it seems that third quantized of gravity with $\phi^4$ interaction term  would naturally 
 lead to the formation of black holes.
In a similar way  $\lambda^2$ processes in the $\phi^3$ theory would naturally lead to the formation of wormholes.  We could also discuss virtual 
processes which give rise to the spacetime foam, in the third formalism. 
 Thus  $\phi^3$ theory would give rise to virtual wormholes \cite{wh} and $\phi^4$ theory would give rise to the 
virtual black hole \cite{vb}, in the spacetime foam.  The low-energy effects of these virtual wormholes or virtual black holes in our universe 
can be given in terms 
of an  effective interaction Lagrangian density given by 
\begin{equation}
 \mathcal{L}_{eff} = \sum_i \mathcal{L}^i \phi_i,
\end{equation}
where the index $i$ labels the different wormholes or virtual black holes 
and  $\mathcal{L}^i$  is the insertion operator at the nucleating
event. Now the cosmological constant can be shown to vanish by repeating the argument used in Ref. \cite{cm}. 
However in Ref.  \cite{cm} only virtual wormholes were considered,  but in the present third quantized formalism
  virtual black holes would also have the same low energy effect on the cosmological constant. 
However as the  universes with matter or anti-matter asymmetry form in pairs, they will remain entangled to each other.  
 This may generate  an effective value for the the cosmological constant. In fact as time passes the entanglement will 
reduce and so will the value of the cosmological constant. This  might explain 
the hight value of the cosmological constant at the early states  of our universe \cite{c}
and its low value now \cite{cc}.
 
A interesting consequence of modifying the Wheeler-DeWitt equation by adding interactions is that it is possible that the vacuum of the present multiverse is 
not a true vacuum. This can be seen by considering the potential term constructed out of a general interaction term as follows:
\begin{equation}
 V[\phi] = - \frac{ \delta \mathcal{S}_{EI} [\phi]}{ \delta \phi }.
\end{equation}
If this  $V[\phi]$ is metastable then there is a false vacuum, and if our multiverse is formed by perturbations around this false vacuum then it can always 
 tunnel to a true vacuum. The amplitude for this tunnelling will be given by 
\begin{equation}
 Z = \int D\phi \exp - \mathcal{S}_{T}.
\end{equation}
In semi-classical approximation this decay rate of the false vacuum to a true one  will be given by 
\begin{equation}
 \Gamma = A \exp - B, 
\end{equation}
where $B$ is the coefficient of the third quantized Euclidean action evaluated at the bounce and $A$ is a product of certain factors 
and a square root of the absolute value of 
the determinant of the second variation of the third quantized Euclidean action evaluated at the bounce.   
Thus if our multiverse is formed by perturbations around a false third quantized vacuum it can tunnel to a true third quantized vacuum. 
In that case all the structure  in our multiverse would be destroyed. The possibility of our universe being formed around a false second quantized
 vacuum has been already discussed in Ref. \cite{p, q}, where  it was concluded that if our universe is formed around a false second quantized vacuum, then there 
is a finite probability for it to tunnel to a true second quantized vacuum.  
We have applied this idea to the whole multiverse by considering the multiverse to be formed around a
false third quantized vacuum state.  
\section{Conclusion}
In this paper we have obtained the wavefunction for the multiverse by third quantization of the Horava-Lifshits theory of gravity without the detailed
balance condition and shown that it leads to the creation of universes from nothing. Furthermore, we have shown that baryogenesis  without the violation of the 
conservation of the baryon number  occurs due to  the third quantization of a 
 modified Wheeler-DeWitt equation.  We also discuss
a possible solution to the cosmological constant problem and the consequences of a third quantized false vacuum.
 
The advantage of using third quantization is that it naturally describes the multiverse. Many results that have been discussed for a single universe in the 
second quantized formalism, can be easily generalized  to the full multiverse in the third quantized formalism.  
Another advantage of using third quantization is that 
the potential of second quantized theories becomes the frequency for third quantized theories. It is more convenient to deal with frequency than with potentials. 

An important assumption made here is that the scalar factor acts like the time variable \cite{l1}. This can only be done in  minisuperspace models,
it is not clear how these results can be generalized  to the full superspace and what can act as time in the full superspace. It is possible to write 
the Wheeler-DeWitt equation in full superspace as a time-independent Schroedinger wave equation with  cosmological constant as its eigenvalue \cite{7r, 8r}. 
Thus if we are able to write a  time-dependent version of the Wheeler-DeWitt equation in analogy to time-dependent Schroedinger wave equation, which would reduce 
to the conventional Wheeler-DeWitt equation for states of fixed cosmological constant,
 then we could  obtain   time on the full superspace.   
After which  we can apply the  methods developed here for minisuperspace to the full superspace.

\end{document}